\documentclass[aps,prl,twocolumn,superscriptaddress]{revtex4}
\usepackage{graphicx}

\newcommand{\omegaL}{\omega_\mathrm{L}}

\newcommand{\omegaS}{\omega_\mathrm{S}}

\newcommand{\TE}{\tau_\mathrm{E}}
\newcommand{\ELO}{E_\mathrm{LO}}
\newcommand{\EO}{E}
\newcommand{\EL}{E_\mathrm{L}}
\newcommand{\EI}{E_\mathrm{I}}

\newcommand{\lambdaL}{\lambda_{\rm L}}

\begin{document}


\title{Parallel heterodyne detection of dynamic light scattering spectra from gold
nanoparticles diffusing in viscous fluids.}

\author{Michael Atlan}
\affiliation{Institut Langevin, CNRS UMR 7587, INSERM U 979, Fondation Pierre-Gilles de Gennes, Universit\'es Paris 6 \& 7, ESPCI ParisTech, 10 rue Vauquelin, 75005 Paris, France.}

\author{Pierre Desbiolles}
\author{Michel Gross}
\affiliation{Laboratoire Kastler-Brossel de l'\'Ecole Normale
Sup\'erieure, CNRS UMR 8552, Universit\'e Pierre et Marie Curie -
Paris 6, 24 rue Lhomond 75231 Paris cedex 05. France}

\author{Ma\"it\'e Coppey-Moisan}
\affiliation{D\'epartement de Biologie Cellulaire, Institut Jacques
Monod, CNRS UMR 7592, Universit\'e Paris 7, 2 Place Jussieu,
Tour 43, 75251 Paris Cedex 05. France}

\date{\today}

\begin{abstract}
We developed a microscope intended to probe, using a parallel
heterodyne receiver, the fluctuation spectrum of light
quasi-elastically scattered by gold nanoparticles diffusing in viscous fluids. The cutoff frequencies of the recorded spectra scale up linearly with those expected from single scattering formalism in a wide range of dynamic viscosities (1 to 15 times water viscosity at room temperature). Our scheme enables ensemble-averaged optical fluctuations measurements over multispeckle recordings in low light, at temporal frequencies up to 10 kHz, with a 12 Hz framerate array detector. OCIS : 180.6900, 090.1995, 290.5820
\end{abstract}

\maketitle

Advances in dynamic light scattering microscopy  \cite{Maeda1972,
Nishio1985, Blank1987, BarZiv1997} have recently raised
the possibility of developing passive microrheology tools based on
the measurement of local thermally-driven intensity fluctuations
\cite{Kaplan1999, MasonGangWeitz1996}. The related homodyne
measurement  schemes involve usually single-mode detection
\cite{Brown1987, Ricka1993, Gisler1995}. Concurrently, heterodyne detection is a powerful method for investigating dynamic light scattering (DLS) in low-light
conditions \cite{BernePecora2000, HohlerCohenAddad2003}. In usual heterodyne
detection schemes, the intensity fluctuations of the mix between the
scattered field and a local oscillator (LO) field are measured on
a single-mode detector. Since the probed scattered
field matches the LO beam mode structure \cite{Cohen1975}, such heterodyne schemes cannot perform multimode measurements. Multimode homodyne techniques based on intensity fluctuations measurements have been developed to take advantage of the spatial bandwitdh of array detectors in order to perform speckle ensemble
averaging and/or spatially-resolved measurements \cite{Wong1993,
Kirsch1996, Cipelletti1999, ViasnoffLequeux2002,
ScheffoldCerbino2007}. However, these parallel homodyne techniques suffer from a temporal frequency bandwidth imposed by the camera frame rate and from limited sensitivity in low light.

We present here a DLS microscope designed for probing the
first-order Doppler spectrum of the light scattered by
nanoparticles in Brownian motion. Our microscope performs heterodyne detection with
a camera used as a multi pixel detector. The Doppler spectrum is
explored by sweeping the LO beam optical frequency, so that the
temporal frequency bandwidth is not limited by the camera frame
rate.

\begin{figure}[]
\centering
\includegraphics[width = 6 cm]{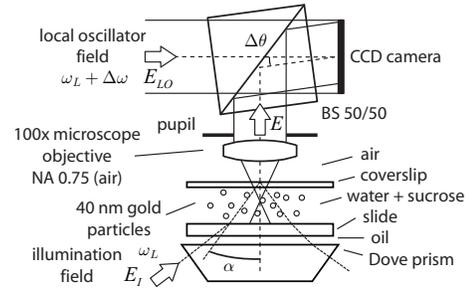}
\caption{Optical configuration. A colloidal gold  suspension in a
calibrated viscous fluid is illuminated by single mode laser
light. The total internal reflection at the top interface ensures
a dark-field illumination of the particles. The light field $\EO$
scattered by the sample beats against a local oscillator field
$\ELO$, detuned by $\Delta \omega$ with respect to the
illumination field $\EI$. $\ELO$ is tilted by $\Delta \theta$ with
respect to $\EO$. The interference pattern $\left| \EO + \ELO
\right|^2$ is recorded by a CCD camera.}\label{fig_setup}
\end{figure}

The optical configuration is sketched  in Fig.\ref{fig_setup}. The
whole optical apparatus has been described elsewhere
\cite{AtlanGrossDesbiolles2008}. A laser beam ($\lambdaL$ = 532
nm, field $\EL$, angular frequency $\omegaL$, average power 50 mW,
single axial mode, CW) is split into tunable-frequency local oscillator (LO) and illumination arms (fields $\ELO$ and $\EI$) to form a Mach-Zehnder
interferometer. The samples are thin chambers
(thickness $h \sim 100 \, \mu \rm m$, several millimeters wide)
containing an aqueous solution of gold particles (diameter $d$ = 40 nm :
BBI international, EM.GC40) with added sucrose to
increase the viscosity. The particle density is $\rho \sim 2.0
\times 10^{16} \rm \, m^{-3}$. At $\lambdaL = 535 \, \rm nm$, the
integral scattering cross-section $\sigma_s$ of a 40 nm gold
particle normalized by the particle volume $v$ is $\sigma_s / v
\sim 5.2 \, \mu {\rm m} ^{-1}$ \cite{Jain2006}. Since the optical mean free path is $l = (\rho \sigma_s)^{-1}$, the ratio of the sample thickness to the mean free path is thus about $h/l \sim 3.5
\times 10^{-4}$, optical fluctuations are interpreted with single scattering formalism in dilute suspensions \cite{BernePecora2000}. A Dove prism guides the illumination field $\EI$ to the sample. Microscope immersion oil between the prism and the slide enables refractive index matching. To prevent the collection of the unscattered field, a total internal reflection is provoked at the glass-air interface on top of the sample. The incident field makes an angle $\alpha \sim 50^\circ$ in water with respect to the optical axis. The field scattered by the gold beads $\EO \ll \ELO$ is collected by a microscope objective (Leica PL Fluotar: 100 $\times$ magnification, $\rm NA = 0.75$, air, exit pupil diameter $\simeq 3$ mm). The LO linear polarization is tuned to match the dominant polarization axis of the object field. Off-axis optical mixing ($\Delta \theta=0.08$ rad tilt) of $\EO$ and $\ELO$ with a
beam splitter (BS) results in an hologram $I = \left| \EO + \ELO
\right|^2$ that is recorded with a CCD camera (PCO Pixelfly QE,
$1392 \times 1024$ pixels of $d = 6.45 \, \mu \rm m$, frame rate
$\omegaS/(2\pi) = 12 \, \rm Hz$, exposure time $\TE$ = 83 ms) from
which only the central $N_x \times N_y = 1024 \times 1024$ pixels region is used
for the calculations.


\begin{figure}[]
\centering
\includegraphics[width = 8 cm]{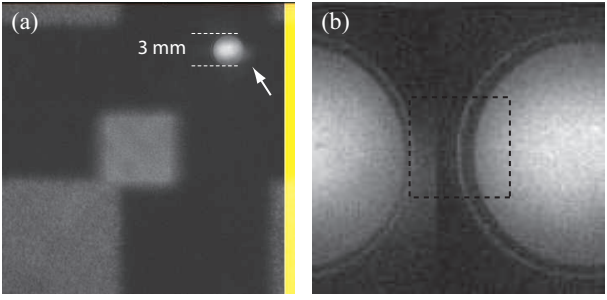}
\caption{(a) Hologram ($1024 \times 1024$ pixels) calculated in
the exit plane of the objective (arrow). (b) Zoom ($141 \times
141$ pixels) of the objective pupil.
}\label{fig_pplane}
\end{figure}

The DLS spectra are acquired by sweeping the detuning frequency
$\Delta \omega$ of the LO. At each frequency point $\Delta \omega$, 16
successive CCD frames $\{I_1, ..., I_{16}\}$ are recorded. The
field component $\EO$ in the detector plane is
calculated from the difference of consecutive images
$I_{n+1}-I_{n}$. This
quantity is back-propagated numerically up to the exit plane of the
microscope objective by a standard convolution method
\cite{SchnarsJuptner1994} with a reconstruction distance $D= 34.9$
cm. A representation of the squared amplitude of the
back-propagated hologram is shown in Fig.\ref{fig_pplane} (a).
Because of the angular tilt $\Delta \theta$ between the interfering beams, the
image of the objective is shifted in the upper right corner of
the hologram. Fig.\ref{fig_pplane} (b) shows a zoom of the
objective pupil image. The heterodyne signal $s_{1} \propto \left| \EO \right| ^2$ is averaged within the square region shown in Fig.\ref{fig_pplane} (48 $\times$ 48 pixels with pixel size $\lambda  D / (N_x d) = 28 \, \mu \rm m$).

In the time domain, the normalized autocorrelation function
$g_{1d}$ of the fluctuating object field in the detector plane
is \cite{BernePecora2000} :
\begin{equation}\label{eq_g1d}
g_{1d}(q,\tau) = \exp(-D q^2 \tau) = \exp(- \tau/\tau_c),
\end{equation}
where $\tau_c$ is the relaxation time, $q$ is the scattering
vector (momentum transfer), and $D$ is the particle diffusivity. Since the beads undergo Brownian motion, the average mean-square displacement of
each particle at time $t$ scales as $\sim D t$, where the
diffusion constant $D$ is linked to the thermal energy $k_B T$ and
to the mobility $\mu$ of the particles: $D = \mu k_B T$. In the
limit of low Reynolds number $\mu = 1/(6 \pi \, \eta \, r)$ ,
where $\eta$ is the viscosity of the medium and $r$ is the
(hydrodynamic) radius of the spherical particles. The illumination and collection geometry introduce momentum transfer diversity, and because we seek to study dynamic fluctuations in non-varying optical scattering conditions, we retain
\begin{equation}\label{eq_g1d_approx}
g_{1d}(\tau) \approx \exp (-D a k^2 \tau)
\end{equation}
where $k = 2 \pi / \lambdaL$ is the angular wavenumber and $a$ is an adjustment parameter used to describe the scattering structure factor of the optical configuration. Additionally, the electric field $\EO$ that reaches the detector is a superposition of a field $E_s$ statically scattered and a field $E_d$ dynamically scattered by the sample \cite{BoasYodh1997, ZakharovVolker2006}. The normalized autocorrelation function $g_{1}$  of the total field $\EO = E_d + E_s$ is
\cite{BoasYodh1997} :
\begin{equation}\label{eq_g1}
g_1(\tau) = (1-\rho) \left|g_{1d} (\tau)  \right| + \rho,
\end{equation}
where $\rho = \left| E_s \right|^2 /  (\left| E_d \right|^2 +
\left| E_s \right|^2)$.
%
%
According to the Wiener-Khinchin theorem, the first-order power
spectral density $s_1 (\omega)$ of the field fluctuations
is the Fourier transform of the field autocorrelation function
$g_1(\tau)$. Let $s_{1d} (\omega)$ be the Fourier transform of
$\left| g_{1d} (\tau) \right|$. Taking into account the instrumental response $B$ in the frequency domain, we have :
\begin{equation}\label{eq_s1}
s_1(\omega) = (1-\rho) \, s_{1d}(\omega) * B(\omega) + \rho \, B(\omega),
\end{equation}
where $*$ is the convolution product. In the absence of dynamic scattering, the off-axis component of the recorded image $I_n$ at time $t_n = 2 n \pi/\omegaS$ is the interference term $\EO \ELO^* \exp (i \Delta \omega t_n)$. The measured signal $I_n ^{\rm off-axis}$, integrated over an exposure time $\TE$ is :

\begin{equation}\label{eq_In}
I_n ^{\rm off-axis} =  \frac{1}{\TE} \int _{0} ^{\TE} \EO \ELO^* \exp [i \Delta \omega ( t_n + \tau )] \, {\rm d}\tau
\end{equation}

where $i^2=-1$. The normalized apparatus lineshape $B(\Delta \omega) = | I_{n+1}^{\rm off-axis} - I_n^{\rm off-axis} |^2 / (16 | \EO |^2 | \ELO |^2 )$ is then :
\begin{equation}\label{eq_B}
B(\Delta \omega) = (\Delta \omega \TE)^{-2} \sin ^2 (\Delta \omega \TE /2) \, \sin ^2 (
\pi \Delta \omega / \omegaS).
\end{equation}

\begin{figure}[]
\centering
\includegraphics[width = 6.5cm]{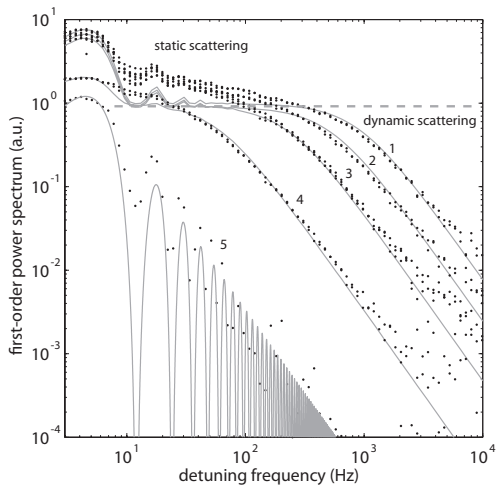}
\caption{First-order light fluctuations spectra normalized by the
relative weight of dynamically scattered light.  The relative viscosities of the samples \cite{Lide2004} are $\eta/\eta_0$ = 1.00 (1), 1.69 (2), 3.94 (3), and 15.40 (4). Continuous lines are plotted from Eq.\ref{eq_s1}. Data points (5) represent the apparatus lineshape in the frequency domain, and the continuous line is
plotted from Eq.\ref{eq_B}. The horizontal axis is the detuning frequency $\Delta \omega/(2\pi)$.}\label{fig_080731nanoDLS1}
\end{figure}

This function is plotted on fig. \ref{fig_080731nanoDLS1} (line 5). Hence, in good approximation, frequency shifts $\Delta \omega$ probe actual fluctuations frequencies $\omega$. The normalized first-order power spectrum of the fluctuating field is the Fourier transform of equation \ref{eq_g1d} :
\begin{equation}\label{eq_s1d}
s_{1d} (\omega) = 1 / (1 + \omega^2 \tau_c^2),
\end{equation}
where $\tau_c = 6 \pi \eta r / (q^2 k_B T)$. Experimentally, the
samples must be prepared with great care to avoid stray light. Otherwise, the estimation of the cutoff frequency $\tau_c ^{-1}$ is made difficult because both fluctuating and non fluctuating light are recorded within the available CCD dynamic range. A high relative weight of static light prevents an accurate detection at high frequencies because it reduces the effective dynamic range.

The measured spectra from samples of relative viscosity
$\eta/\eta_0$ = 1.00 (1), 1.69 (2), 3.94 (3), and 15.40 (4) are
represented in Fig.\ref{fig_080731nanoDLS1}, where $\eta_0 =
10^{-3} \, \rm Pa.s$ is pure water viscosity. For each
measurement, the frequency is detuned from 3 Hz to $10^4$ Hz over
80 points with logarithmic spacing. Each scan lasts $\sim 106 \,
\rm s$ and 5 minutes are needed to save the data. The
frequency scans are performed twice with 15 minutes intervals
and signal variability from one measurement to another is visible
at $10^{-3}$ from the maximum typically (points clouds). The
$s_{1}$ lines plotted in Fig.\ref{fig_080731nanoDLS1} are calculated from Eq.\ref{eq_s1}, with $k_B T = 4.11 \times 10^{-21} \, \rm J$ at room temperature (298 K), $r = 40/2 = 20 \, \rm nm$, and $q^2 = 3.7 k^2$. The adjustment parameter $a = 3.7$ has been chosen to get a good agreement with the experimental data points. The expected momentum transfer from particle scattering of the incident and reflected beams, in DLS conditions \cite{BernePecora2000} is $q^2 = 4 n_{\rm w}^2 \sin ^2 (\beta/2) \, k^2$. $n_{\rm w} = 1.33$ is the refractive index of water. For the incident beam, we have ${q} ^2 = 1.3 \, k^2$ ($\beta = \alpha = 50 ^\circ$). For the reflected beam, we have ${q} ^2 = 5.8 \, k^2$ ($\beta = \pi - \alpha = 180-50 ^\circ$). The derived cutoff frequencies are ${\tau_c}^{-1}$ = 888 Hz (1), 524 Hz (2), 225 Hz (3), 58 Hz (4); these cutoffs are inversely proportional to the viscosity and spread over a wide frequency range. The values of the relative weight of the dynamic light component for the different preparations used to fit $s_{1} (\omega)$ to the experimental data are $1-\rho$ = 0.48 (1), 0.15 (2), 0.14 (3), 0.13 (4), 0.0 (5). Data points (5) of Fig.\ref{fig_080731nanoDLS1} are obtained from a preparation without Brownian particles; they are in agreement with the apparatus lineshape derived from Eq.\ref{eq_B}.

These results demonstrate the suitability of the proposed method for
parallel heterodyne dynamic light scattering measurements of nanoparticles dynamics in microscopic volumes. The resolution, range and sensitivity of first-order spectrum measurements potentially enable a robust exploration of a wide range of viscosity-dependent fluctuations at microscopic scales.

The authors acknowledge support from Agence Nationale de la Recherche (ANR-09-JCJC-0113 grant), Fondation Pierre-Gilles de Gennes (FPGG014 grant), R\'egion Ile-de-France (C'Nano grant) and CNRS.


\end{document}